\documentclass[pre,aps,amsmath,amsfonts,amssymb,showpacs,twocolumn]{revtex4-1}
\usepackage{graphicx,bm}
\usepackage{CJK,amscd}
\newcommand{\cM}{{\cal M}}
\begin{document} 
\begin{CJK*}{UTF8}{mj}
\title{Order-disorder transition in a model with two symmetric absorbing states}
\author{Su-Chan Park (박수찬)}
\affiliation{Department of Physics, The Catholic University of Korea, Bucheon 420-743, Korea}
\date{\today}
\begin{abstract} 
We study a model of two-dimensional interacting monomers  
which has two symmetric absorbing states and exhibits two kinds
of phase transition; one is an order-disorder transition and
the other is an absorbing phase transition. Our focus is around the 
order-disorder transition, and we investigate whether this transition is 
described by the critical exponents of the two-dimensional Ising model. 
By analyzing the relaxation dynamics
of ``staggered magnetization,'' the finite-size scaling, and the 
behavior of the magnetization in the presence of a symmetry-breaking field,
we show that this model should belong to the Ising universality class. 
Our results along with the universality hypothesis support the idea
that the order-disorder
transition in two-dimensional models with two symmetric absorbing states
is of the Ising universality class, contrary to the
recent claim [K. Nam {\it et al.}, J. Stat. Mech.: Theory Exp. (2011) L06001].
Furthermore, we illustrate that the Binder cumulant could be a misleading guide
to the critical point in these systems.
\end{abstract}
\maketitle
\end{CJK*}
\section{introduction}
During recent decades, intense theoretical efforts have been devoted to 
classifying absorbing phase 
transitions (APTs) and, as a result, several universality classes have been found
(for a review, see, e.g., Refs.~\cite{MD1999,H2000,O2004,L2004}).
Although a firmly established classification principle is still desired, 
symmetry is unequivocally expected to play an important 
role in determining universality classes. 
On the one hand, if a model which has a single absorbing state
and which does not have extra symmetry or conservation laws
exhibits an APT,
it is argued that this model should belong to the directed percolation (DP)
universality class~\cite{J1981,G1982}. On the other,
many systems with two symmetric (sets of) absorbing
states are known to form  
another universality class. Such examples with $Z_2$ symmetry are
the probabilistic cellular automata model~\cite{GKvdT1984}, the interacting
monomer-dimer (IMD) model~\cite{KP1994}, the nonequilibrium kinetic 
Ising model~\cite{M1994}, and the interacting monomer-monomer 
model with infinitely many absorbing states~\cite{PP2001}, to cite only a few. 

Although symmetry seems important, 
there are also some systems with $Z_2$ symmetry
which do not share criticality with the above-mentioned models.
Such exceptions can be found in Ref.~\cite{PP2009}. Hence,
symmetry alone is not sufficient to determine the universality class
and further studies are necessary to determine a guiding principle in terms of 
symmetry.

The starting point to develop a principle would be construction
of a (coarse-grained) field theory for each universality class,
followed by a renormalization group analysis.
The connection between DP and the reggeon field theory has 
been clarified long time ago~\cite{CS1980} (see also \cite{J1981,G1982}).
Endeavors have also been made to formulate a field theory for 
systems with $Z_2$-symmetric absorbing states.
Cardy and T\"auber~\cite{CT1996,CT1998} developed the field theory for
the branching annihilating random walks with even numbers of 
offspring (BAWE)~\cite{TT1992}, which has mod-2 conservation of particle 
number. Since  $Z_2$ symmetry entails the mod-2 
conservation of domain walls in one dimension, the BAWE in one dimension
belongs to the same universality class as systems with two symmetric
absorbing states.
The mod-2 conservation, however, has nothing to do with the $Z_2$ symmetry
in higher dimensions. 
In particular, the BAWE in higher dimensions
shows trivial transitions~\cite{CT1996,CT1998}, which
is not the case for two-dimensional models with 
$Z_2$ symmetry~\cite{DFL2003,HCDM2005,VL2008,NPKL2011}.
In this regard, the field theory of the BAWE cannot be a coarse-grained 
description for models with $Z_2$ symmetry in $d$ dimensions.
Thus, there was a theoretical request to develop a field theory 
with $Z_2$ symmetry in higher dimensions 
and, as a response,  a phenomenological Langevin equation was 
introduced~\cite{HCDM2005} (see also Ref.~\cite{CCDDM2005} for an analysis
of the corresponding field theory by the nonperturbative renormalization group method).

Recently, however, this phenomenological Langevin equation description has  
been challenged~\cite{NPKL2011}.
The Langevin equation predicts that there are in general two transitions
in two dimensions; an order-disorder transition which is concomitant with the 
$Z_2$ symmetry breaking (SB) and an APT
(see Ref.~\cite{DFL2003} for the first observation of two transitions
in a microscopic model). Numerical analysis of the Langevin equation 
revealed~\cite{HCDM2005} that the $Z_2$ SB transition is of the 
Ising class and the APT is of the DP class.
Although a Monte Carlo simulation of the two-dimensional IMD model also
found a $Z_2$ SB transition followed by an APT, 
it was claimed that the critical behavior for the $Z_2$ SB transition 
is not of the Ising class~\cite{NPKL2011}.
In fact, no Monte Carlo simulations studies up to now, to our knowledge,
have clearly shown that the $Z_2$ SB occurring in a system with two symmetric
absorbing states is described by the Ising critical exponents, which was
the motivation of Ref.~\cite{NPKL2011}. If the claim 
in Ref.~\cite{NPKL2011} turns out to be true, a different coarse-grained 
description from that suggested is called for. Even more seriously, 
the conclusion in Ref.~\cite{NPKL2011} questions the validity of 
the theory that any continuous $Z_2$ SB transition between
ordered and disordered phases should be described 
by the scalar $\phi^4$ theory~\cite{GJH1985}, 
which was the motivation to introduce 
the model-$A$ type (according to the Hohenberg-Halperin classification 
scheme~\cite{HH1977}) interaction to the Langevin equation~\cite{HCDM2005}.

Hence, it is necessary to study the $Z_2$ SB transition exhibited
by a two-dimensional model with two symmetric absorbing states 
more extensively to 
make a firm conclusion concerning the universality class.
In this paper, we thoroughly investigate the SB transition.
The model studied in this paper will be called the two dimensional
interacting monomers (2DIM)  model which is a two-dimensional
version of the model studied in Ref.~\cite{PP2008PRE}.

The paper is organized as follows: After introducing
a model and appropriate order parameters in Sec.~\ref{Sec:model},
we present numerical analysis of the SB transition
in Sec.~\ref{Sec:num}. 
Section~\ref{Sec:dis} discusses 
the claim in Ref.~\cite{NPKL2011}, studies the absorbing 
phase transition to confirm the DP transition, and then summarizes the paper.

\section{\label{Sec:model}Model}
The 2DIM model is defined on a square 
lattice with size $L^2$ ($L$ is assumed to be even). Every site is indexed 
by a two dimensional vector $\bm{x} = (i,j)$ with integer components 
$i$ and $j$ ($i,j=0,1,\ldots, L-1$). 
Periodic boundary conditions are assumed.
For later purposes, the lattice is subdivided into two sublattices $E$ and $O$.
The sublattice $E$ ($O$) is defined as a set of sites 
$\bm{x}= (i,j)$ with $i+j$ even (odd). Each site is either occupied by 
a monomer or vacant. Two monomers are not allowed to occupy a single site
at the same time.
Each site is given a state variable $a_{\bm{x}}$
which takes the value 1 (0) if site ${\bm{x}}$ is occupied (vacant).
A configuration is characterized by state variables at all sites.

The dynamic rules are as follows: A monomer attempts to adsorb on a 
randomly chosen vacant site (called a target site). 
Depending on the number $n$ of occupied nearest neighbors of the target
site, the fate of the monomer will be different.
If $n=0$, the monomer adsorbs with 
rate 1. On the other hand, if $n\neq 0$, the monomer 
adsorbs with rate $n \lambda_n$, but the adsorbed monomer 
immediately forms a dimer with a randomly chosen monomer among $n$ 
monomers on the nearest neighbor sites,
and the dimer is desorbed in no time. Effectively, an adsorption event
on a vacant site with occupied nearest neighbors removes one monomer from
the lattice. If all nearest neighbors of the target site are occupied,  that is,
if $n=4$, a monomer is not allowed to adsorb on the target site, 
which amounts to setting $\lambda_4$ to zero.
Since $\lambda_4 = 0$, any configuration with all vacant sites surrounded
by monomers is an absorbing state.

To write the master equation for the above dynamics in a succinct way, 
we introduce a mathematical notation; 
for any configuration $C$ with the state variable
$a_{\bm{y}}$ for every site $\bm{y}$, $C_{\bm{x}}$ stands for the configuration 
obtained by changing the state variable at site $\bm{x}$ 
to $1 - a_{\bm{x}}$ and by keeping all other state variables the same as in $C$. 
Using this notation, the master equation can be written as
\begin{equation}
\frac{d}{dt}P(C;t)
= \sum_{\bm{x}} \left [
W_{CC_{\bm{x}}} P(C_{\bm{x}};t)
-W_{C_{\bm{x}}C}P(C;t)
\right ],
\label{Eq:master}
\end{equation}
with the transition rate
\begin{eqnarray}
W_{C_{\bm x}C} = \delta_{a_{\bm{x}},0} \delta_{n_{\bm{x}},0}+
\delta_{a_{\bm{x}},1} \sum_{\bm{y}}{}' \sum_{k=1}^3
 \lambda_k \delta_{a_{\bm{y}},0}\delta_{n_{\bm{y}},k},
\end{eqnarray}
where $\sum{}'$ means the sum over 
nearest neighbor vectors $\bm{y}$ of site $\bm{x}$, $a_{\bm{x}}$ 
($a_{\bm{y}}$) is the state variable at site $\bm{x}$  ($\bm{y}$) in the 
configuration $C$, $n_{\bm{x}}$ ($n_{\bm{y}}$) means the number of 
occupied nearest neighbors of site $\bm{x}$ ($\bm{y}$) in $C$, 
 and $\delta$ is the Kronecker
delta symbol. By observing that $(C_{\bm{x}})_{\bm{x}} = C$, one can easily
find the transition rate $W_{CC_{\bm{x}}}$.

To simulate the master equation, we have used the 
following algorithm.  First, we make a list of
vacant sites with at least one vacant nearest neighbor.
For convenience, we will refer to such a vacant site as an active site.
Assume that there are $N_t$ active sites at time $t$. 
A target site out of $N_t$ active sites is selected at random with 
equal probability. If all nearest neighbors of the target site are vacant, 
it becomes occupied with probability $\Delta t$
which is defined as
\begin{equation}
\Delta t = \frac{1}{\textrm{max}(1,\lambda_1,2\lambda_2,3\lambda_3)}.
\end{equation}
If the target site has $n$ occupied nearest
neighbors ($n=1$, 2, or 3), a configuration change can occur
with probability $n \lambda_n \Delta t$.
If a change is destined, one monomer out of $n$ is selected with
equal probability and it is removed from the system, which 
mimics the dimer desorption explained above.
After the above attempt, time increases by $\Delta t/N_t$ and 
the list of active sites is updated in an appropriate way.
We repeat the above procedure until
$t$ exceeds a preassigned maximum observation time or 
no active site exists in the system. 
For convenience, we set $\lambda_1 = 2 \lambda_2 = 3 \lambda_3 = \lambda$
in what follows and study phase transitions by tuning $\lambda$.

Now we will specify the initial condition studied in this paper.
At $t=0$, the sublattice $O$ is empty, but a site in the sublattice $E$
is occupied with probability $m_0$ ($0\le m_0 < 1$).
With this initial condition, there are only two absorbing states;
the sublattice $E$ is fully occupied and the  sublattice $O$
is empty, or vice versa.  In this sense, absorbing states 
have perfect `anti-ferromagnetic' order.
The absorbing state with the sublattice $E$ ($O$) filled with monomers 
will be called the even (odd) absorbing state. 

Since we expect two transitions (an APT and
a $Z_2$ SB transition), two quantities that are respectively called 
the density of active sites and the `staggered magnetization' 
will be measured during simulations.

The density of active sites at time $t$ is defined as
\begin{equation}
\phi(t,L) \equiv \frac{1}{L^2}
 \sum_{\bm{x}} \delta_{a_{\bm{x}},0} ( 1 - \delta_{n_{\bm{x}},4} ),
\label{Eq:abs_order}
\end{equation}
where $n_{\bm{x}}$ is the number of occupied nearest neighbors of site $\bm{x}$
in a configuration at time $t$. If the system is in one of the two absorbing states at time $t$,
$\phi(t,L)$ is obviously zero. 
We define the (averaged) density of active sites in the thermodynamic limit as
\begin{equation}
\rho(t) \equiv \lim_{L\rightarrow \infty} \langle \phi(t,L) \rangle,
\end{equation}
where $\langle \ldots \rangle$ stands for the average over all independent
realizations.

The staggered magnetization (SM) is defined as
\begin{equation}
\cM (t,L) \equiv \frac{1}{L^2} \left [ \sum_{\bm{x} \in E} - \sum_{\bm{x} \in O} \right ]
\left [ 2 a_{\bm{x}}(t) - 1 \right ].
\end{equation}
If the system is in the even (odd) absorbing state, the SM is $\cM(t,L) = 1$ ($-1$). 
The (averaged) SM at time $t$ in the thermodynamic limit is 
defined as 
\begin{equation}
m(t) \equiv \lim_{L\rightarrow \infty} \langle \cM(t,L) \rangle.
\end{equation}
\begin{figure}[t]
\includegraphics[width=0.48\textwidth]{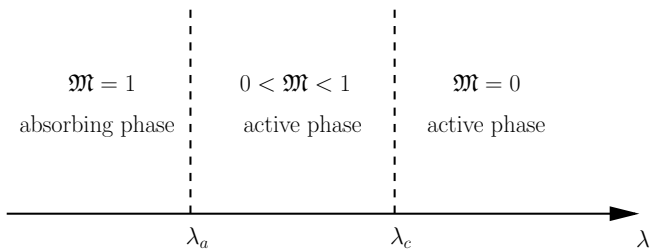}
\caption{\label{Fig:pb} Phase diagram of the 2DIM model. $\lambda_c$ is the transition point of the order-disorder transition and $\lambda_a$ is that of the
absorbing phase transition.
}
\end{figure}

Since the initial condition gives
\begin{eqnarray}
\left \langle a_{\bm{x}} \right \rangle 
= \begin{cases}
0& \text{ if } \bm{x} \in O,\\
m_0& \text{ if } \bm{x} \in E,
\end{cases}
\\
\left \langle \delta_{a_{\bm{x}},0}\left (1- \delta_{n_{\bm{x}},4}\right )\right \rangle
= \begin{cases}
1-m_0^4& \text{ if } \bm{x} \in O,\\
1-m_0& \text{ if } \bm{x} \in E,
\end{cases}
\end{eqnarray}
we get
\begin{equation}
\langle {\cal M}(0,L) \rangle = m_0,\quad
\langle \phi(0,L) \rangle = 1 - \frac{1}{2}m_0 \left (1 +  m_0^3 \right ).
\label{Eq:ini_con}
\end{equation}
Note that with the above initial condition,
$m(t)$, which is defined in the thermodynamic limit, will remain 
positive for all finite $t$ if $0<m_0<1$  
and the order-disorder phase transition is 
defined by the infinite-time limit of $m(t)$ such that
\begin{equation}
\mathfrak{M}\equiv \lim_{t\rightarrow \infty} m(t) = 
\begin{cases}
\text{nonzero,} & \text{ordered phase,}\\
0, & \text{disordered phase.}
\end{cases}
\end{equation}

As we will see later, the 2DIM model exhibits two transitions: an order-disorder
transition occurring at $\lambda = \lambda_c$ and an absorbing phase
transition occurring at $\lambda = \lambda_a <\lambda_c$. The schematic
phase diagram is depicted in Fig.~\ref{Fig:pb}.

\section{\label{Sec:num}Numerical analysis of the order-disorder transition}
In this section, we present the simulation results, focusing on the 
order-disorder transition. Rather than studying the Binder cumulant, 
we analyze how $m(t)$ approaches the steady state value.
This approach is also known as the nonequilibrium relaxation 
method~\cite{OI2007}.

At criticality, the SM is expected to decay as~\cite{S1976,NKL2008}
\begin{equation}
m(t) \sim t^{-\beta/(\nu z)},
\end{equation}
with the critical exponents $\beta$, $\nu$, and $z$ defined as,
\begin{equation}
\mathfrak{M} \sim (\lambda_c - \lambda)^\beta,\quad \xi \sim |\lambda_c - \lambda|^{-\nu},\quad
\tau \sim |\lambda_c - \lambda|^{-\nu z},
\end{equation}
where  $\lambda_c$ is the critical point,
$\xi$ is the correlation length, and $\tau$ is the relaxation time.
For the two dimensional
Ising model, $\beta = \frac{1}{8}$ and $\nu = 1$ are known exactly
(see, for instance, Ref.~\cite{Baxter}), but
$z \simeq 2.17$~\cite{NKL2008} is known only numerically;
nonetheless it serves well for our purpose.
In simulations, we set $m_0 = 0.9$ (initial condition) and observed how $\langle \cM(t,L=2^{11})
\rangle$ behaves up to $t=5\times 10^5$. 
When we study the finite-size scaling later, 
the finite size effect is argued to be negligible up to
the observation time in this case, so $\langle \cM(t,L=2^{11}) \rangle$ can be
 regarded as $m(t)$. 

\begin{figure}[t]
\includegraphics[width=0.48\textwidth]{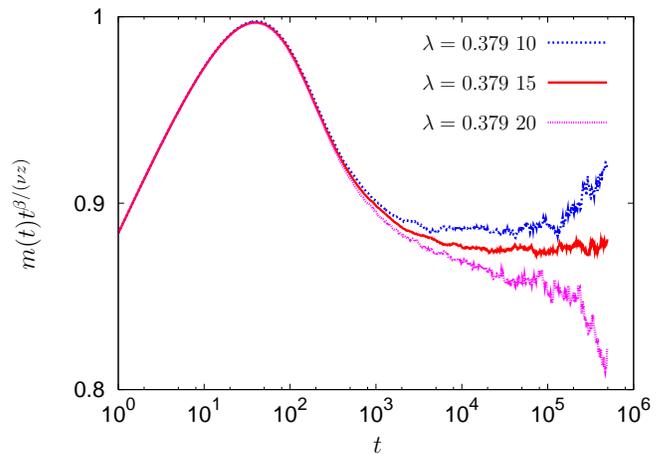}
\caption{\label{Fig:mt} (Color online) Plots of $m(t)
t^{\beta/(\nu z)}$ vs $t$ with Ising critical exponents $\beta = \frac{1}{8}$, 
$\nu =1$, and $z = 2.17$ 
for $\lambda = 0.379~10$, 0.379 15, and 0.379 20 (from top to bottom)
on a semilogarithmic scale.
}
\end{figure}
In Fig.~\ref{Fig:mt}, we depict 
the behavior of $m(t) t^{\beta/(\nu z)}$ for three different values of 
$\lambda$ as a function of $t$ on a semilogarithmic scale, 
using the critical exponents of the two-dimensional Ising model.
The numbers of independent simulation runs for $\lambda =0.3791$, 
0.379 15, and 0.3792 are 280, 912, and 160, respectively.
In the ordered (disordered) phase, the curve is expected to 
veer up (down), and at criticality the curve should be
flat, if the correct exponents are used.
Thus, Fig.~\ref{Fig:mt} supports the idea that the order-disorder transition in 
the 2DIM model is of the Ising type with the critical point $\lambda_c 
= 0.379~15(5)$, where the number in parentheses indicates the error
of the last digit. 

Note that the power-law behavior of $m(t)$ is observable only for $t>10^4$, 
which implies that the 2DIM model has stronger corrections to
scaling than the two-dimensional Ising model (for example, Fig. 1 in  
Ref.~\cite{NKL2008} shows that the two-dimensional Ising model is already in the scaling regime from $t=10$). As we will see later, the strong
corrections to scaling also plague the behavior of the Binder cumulant,
which will be given as the reason why previous studies could not 
successfully report the universal value of the Binder cumulant 
(see Sec.~\ref{Sec:Binder}).

\begin{figure}[t]
\includegraphics[width=0.48\textwidth]{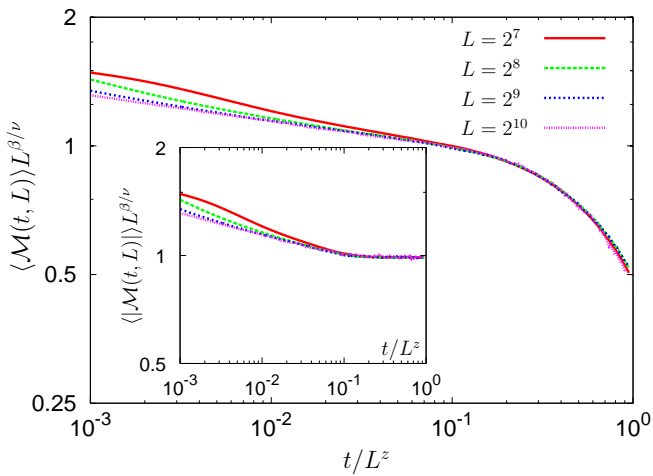}
\caption{\label{Fig:fssm} (Color online)  Plots of $\langle \cM(t,L) \rangle
L^{\beta/\nu}$ vs $t/L^z$ with Ising critical exponents
at criticality on a double logarithmic scale. Inset: Plots of $\langle |\cM(t,L)| \rangle L^{\beta/\nu}$ vs $t/L^z$ at $\lambda =  \lambda_c$.
}
\end{figure}

To have further support for the Ising critical behavior, we also studied 
the finite size scaling. At criticality, scaling collapses
for the magnetization $\cM$
and for the absolute value of the magnetization $|\cM|$ 
are expected with the scaling forms
\begin{eqnarray}
\langle \cM(t,L) \rangle = L^{-\beta/\nu} f(t/L^z),\\
\langle | \cM(t,L) | \rangle = L^{-\beta/\nu} g(t/L^z),
\end{eqnarray}
where $f$ and $g$ are (universal) scaling functions.
To check the finite-size scaling at criticality, we simulated the systems
with sizes of $L=2^7$, $2^8$, $2^9$, and $2^{10}$.
The numbers of independent simulation runs for 
$L=2^7$, $2^8$, $2^9$, and $2^{10}$ are 160 000, 40 000, 10 000, and 2504, 
respectively.
The resulting scaling collapse is presented in Fig.~\ref{Fig:fssm} which 
indeed shows a nice collapse of $\langle \cM(t,L)
\rangle$ onto a single curve when the Ising critical exponents are employed.
The scaling collapse is also nice for the average of the 
absolute value of $\cM$ (inset of Fig.~\ref{Fig:fssm}).

As a by-product of the finite-size scaling,
we can estimate the time after which the finite-size effect becomes 
significant 
to be $t \approx 0.1 L^z$ (about $1.5 \times 10^6$ for $L=2^{11}$)
and, in turn, 
we affirm that the relaxation dynamics of $m(t)$ is faithfully 
presented in Fig.~\ref{Fig:mt}.

Furthermore, we also studied how the symmetry-breaking field $h$ affects
the behavior of the magnetization at criticality. Like the Ising model,
the SM at the critical point is supposed to behave as~\cite{NKL2008}
\begin{equation}
m(t) = h^{1/\delta} H(t h^{\nu z/(\beta \delta)}),
\end{equation}
where $\delta = 15$ for the two-dimensional Ising model~\cite{Baxter} 
and $H$ is a universal scaling function.
It is worth while to investigate a scaling collapse 
of $m(t) h^{-1/\delta}$ plotted as a function of
$t h^{\nu z/(\beta \delta)}$, using the Ising critical exponents.
Note that for the two-dimensional Ising model $\nu z / (\beta \delta) \approx 1.16$.

To introduce the symmetry-breaking field, we follow the idea in 
Ref.~\cite{HKPP1998}. Now the transition rates take the form
\begin{eqnarray}
W_{C_{\bm x}C} =&& \left ( 
1 - \frac{1-(-1)^{i+j}}{2} h \right )\delta_{a_{\bm{x}},0} \delta_{n_{\bm{x}},0}
\nonumber \\&&+
\delta_{a_{\bm{x}},1} \sum_{\bm{y}}{}' \sum_{k=1}^3
 \frac{\lambda}{k}\delta_{a_{\bm{y}},0}\delta_{n_{\bm{y}},k},
\label{Eq:modi}
\end{eqnarray}
where  $i$ and $j$ are components of the lattice vector $\bm{x}$
and $0 < h < 1$.
Recall that we have set $k \lambda_k = \lambda$ for $k=1,2,3$.
By Eq.~\eqref{Eq:modi}, adsorption on the sublattice $E$ 
is more probable than on the sublattice $O$, which  eventually
breaks the symmetry between even and odd absorbing states.
\begin{figure}[t]
\includegraphics[width=0.48\textwidth]{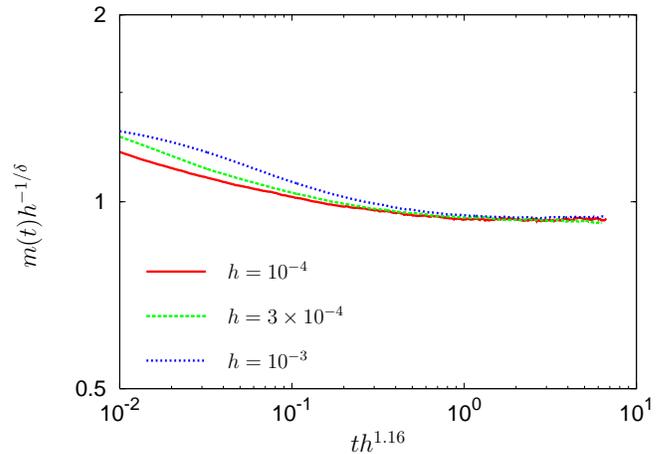}
\caption{\label{Fig:h}(Color online)   Scaling collapse plot of 
$m(t) h^{-1/\delta}$ as a function
of  $th^{1.16}$ on a double logarithmic scale with the Ising critical exponent
$\delta = 15$.
}
\end{figure}

With these modified transition rates, 
we simulated a system with size $L=2^{9}$ at $\lambda =\lambda_c$
for different values of $h$.
In actual simulations, we have only to 
change the probability of adsorption on a vacant site in the sublattice $O$
without occupied nearest neighbors 
to $(1-h)\Delta t$. The resulting scaling plot is depicted in Fig.~\ref{Fig:h}
which shows a nice scaling collapse.
It again supports the idea that the 2DIM model should belong to the Ising universality class.
To make sure that the finite-size effect is negligible, we also studied
systems with size $L=2^8$ and obtained almost same figure as Fig.~\ref{Fig:h} (not shown here).

\section{\label{Sec:dis}Discussion and Summary}
Up to now, we have shown that the critical behavior
of the order-disorder transition in the 2DIM model is of the Ising class. In this section,
we will discuss the behavior of the density of active sites $\rho(t)$ and
the Binder cumulant at the order-disorder transition point $\lambda_c$,
which will be compared with the similar studies in Ref.~\cite{NPKL2011}.
Also, to confirm the universality, we discuss the critical behavior
of the absorbing phase transition.

\subsection{Behavior of $\rho$ and its fluctuation at $\lambda=\lambda_c$}
In Ref.~\cite{NPKL2011}, diverging fluctuation of the order parameter of the absorbing
phase transition was suspected to be a possible reason why the two dimensional
IMD model should not belong to the Ising class. However, this order parameter,
in our case $\phi$ defined in Eq.~\eqref{Eq:abs_order}, seems related to the 
energy density of the Ising model in that $\phi$ is
measured as a correlation between nearest neighbors.
Since the fluctuation
of the (Ising) energy is the specific heat which diverges
logarithmically at criticality in two dimensions, it is actually 
plausible that the fluctuation of $\phi$ (times system size) diverges at
criticality even though the 2DIM model belongs to the Ising class.

\begin{figure}[t]
\includegraphics[width=0.48\textwidth]{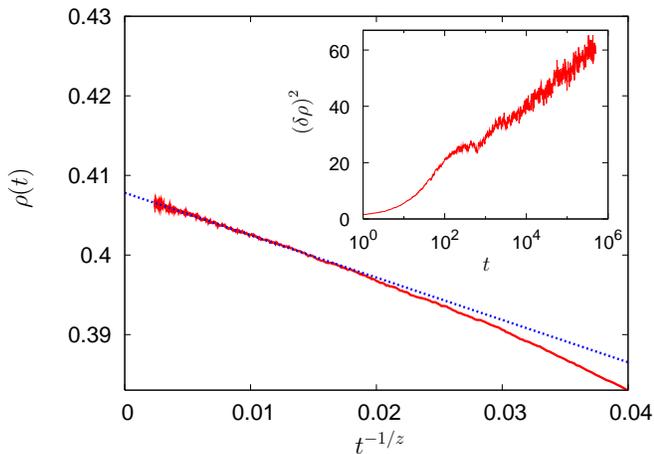}
\caption{\label{Fig:rho_E} (Color online)  Plots of $\rho(t)$ vs $t^{-1/z}$ with the Ising dynamic  exponent $z=2.17$ at criticality. 
The straight line is the result of a linear function fitting.
Inset: Plots of $(\delta \rho)^2$ vs $t$ at
$\lambda = \lambda_c$ on a semilogarithmic scale.}
\end{figure}
To confirm that $\rho(t) = \langle \phi \rangle$ indeed is linked to the 
(average) energy density 
of the Ising model, we analyze how $\rho(t)$ behaves at the order-disorder
transition point. 
Since the energy at criticality approaches the steady-state value
in a power-law fashion with exponent $(\nu d -1)/(\nu z)$~\cite{NKL2008},
$\rho(t)$ is expected, if it is indeed related to the energy, 
to approach the steady-state value $\rho^*$ in such a way that 
\begin{equation}
\rho^* - \rho(t) \sim t^{-(\nu d - 1)/(z \nu)} \approx t^{-1/z},
\end{equation}
where we have set $\nu=1$ (the Ising critical exponent) and $d=2$.
Hence if $\rho(t)$ is plotted as a function of $t^{-1/z}$, the curve becomes
straight for small $t^{-1/z}$ and approaches $\rho^*$ as 
$t^{-1/z} \rightarrow 0$ (equivalently, $t \rightarrow \infty$).
As Fig.~\ref{Fig:rho_E} reveals, $\rho$ approaches the ordinate
as a straight line for $t^{-1/z} < 0.02$, as anticipated.
Note that  the time $t^{-1/z}=0.02$ roughly
corresponds to $t=5\times 10^3$ 
after which $m(t)$ enters the scaling regime (see Fig.~\ref{Fig:mt}).

We also analyzed how the fluctuation of the active site density defined as
\begin{equation}
\left (\delta \rho \right )^2  \equiv \lim_{L\rightarrow \infty} L^2 \left ( \langle \phi(t,L)^2 \rangle - \langle \phi(t,L)
\rangle^2 \right ) 
\end{equation}
behaves at $\lambda = \lambda_c$. 
The inset of Fig.~\ref{Fig:rho_E} shows logarithmic behavior
of $(\delta \rho)^2$ as in the two-dimensional Ising model.

Thus, we conclude that the active site density 
$\phi(t,L)$ is indeed associated with the energy of the Ising model. 
Actually, the logarithmic behavior of $(\delta \rho)^2$ is compatible with the slow divergence
of the fluctuation observed in Ref.~\cite{NPKL2011}.

\subsection{\label{Sec:Binder}Binder cumulant}
\begin{figure}[t]
\includegraphics[width=0.48\textwidth]{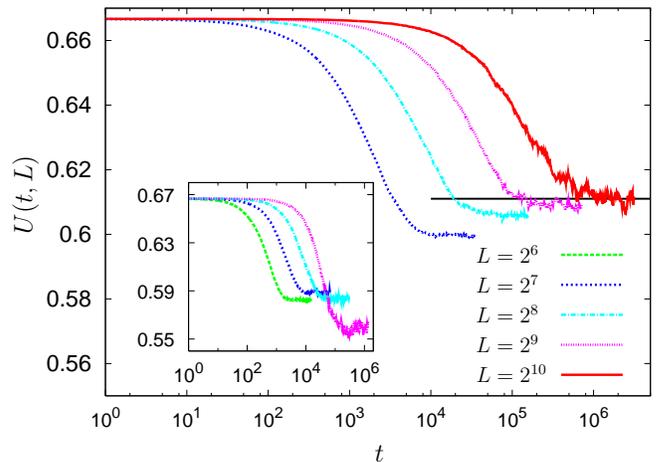}
\caption{\label{Fig:binder} (Color online)  Plots of $U(t,L)$ vs $t$ at criticality
($\lambda = \lambda_c$) on a semilogarithmic scale ($L=2^7$, $2^8$, $2^9$, and $2^{10}$ from left to right). As a guide to the 
eyes, we also plot a straight line indicating the universal Binder cumulant
of the Ising universality class, 0.611. Inset: Plots of $U(t,L)$ vs $t$
in the disordered phase (at $\lambda= 0.3795> \lambda_c$) for $L=2^6$, $2^7$, $2^8$, and $2^9$ from left to right.}
\end{figure}
Since the Binder cumulant is believed to take a universal number at criticality,
we should study whether the Binder cumulant at $\lambda = \lambda_c$ 
approaches to the universal value as $L\rightarrow \infty$.
Defining the Binder cumulant at time $t$ as
\begin{equation}
U(t,L) = 1 - \frac{\left \langle \cM(t,L)^4 \right \rangle}{3 \left \langle 
\cM(t,L)^2 \right \rangle^2},
\end{equation}
we numerically study how $U(t,L)$ behaves for different values of $L$.

In Fig.~\ref{Fig:binder} we present simulation results for $U(t,L)$ for
$\lambda= \lambda_c = 0.379~15$ and for $\lambda = 0.3795 > \lambda_c$ (Inset).
For $\lambda = 0.3795$, the numbers of independent samples simulated for 
$L=2^6$, $2^7$, $2^8$, and $2^{9}$ are 200 000, 50000, 14000, and 4000, 
respectively and data for $\lambda = \lambda_c$ were collected while we
studied the finite size scaling.

Since the system has absorbing states and any finite system will eventually
fall into one of the absorbing states even in the active phase, there are obviously 
two characteristic time scales. 
One is $\tau_q$ when the system enters the quasi-stationary
state and the other is $\tau_a$ when the system falls into one of the absorbing 
states. At $\lambda=\lambda_c$, $\tau_q$ diverges with 
system size as $\tau_q \sim L^z$, but $\tau_a$ should
increase exponentially with $L$ because the SB transition point
is in the active phase of the absorbing phase transition. Hence, 
to find the universal value of the
Binder cumulant at the SB transition point, 
the observation time  should be larger than $\tau_q$ but much smaller
than $\tau_a$.
Actually, except for the case of $L=2^6$, no simulation results in
an absorbing state 
and, even for $L=2^6$, only $\le 0.2\%$ of simulation runs falls
into an absorbing state up to the observation time. Hence, 
in our analysis, the Binder cumulant is not influenced by the existence of 
absorbing states.

\begin{figure}[t]
\includegraphics[width=0.48\textwidth]{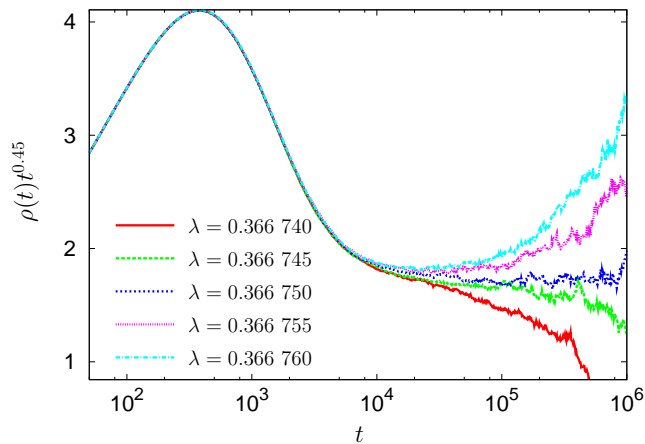}
\caption{\label{Fig:dp} (Color online)  Plots of $\rho(t) t^{0.45}$ vs $t$ near
the critical point of the absorbing phase transition 
on a semilogarithmic scale.}
\end{figure}
At $\lambda = \lambda_c$, $U(t,L)$ in the (quasi-)stationary state increase 
with system size 
but shows a clear signature of saturating behavior to the universal 
number 0.611~\cite{KB1993} as $L \rightarrow \infty$. 
Note that 
if the system size is not large enough, the Binder
cumulant could miidetify the critical point.
The unexpected behavior of the Binder cumulant should be attributed
to the strong corrections to scaling already observed in Fig.~\ref{Fig:mt}.

The inset of Fig.~\ref{Fig:binder} depicts the behavior of $U(t,L)$ 
at $\lambda = 0.3795 > \lambda_c$ (disordered phase).
If the system size is not larger than $L = 2^8$, one may conclude that
the critical point is around 0.3795, with the value of the Binder cumulant 
around 0.59, which is comparable to the value reported in Ref.~\cite{NPKL2011}. 
Hence we conclude that the critical point reported 
in Ref.~\cite{NPKL2011} is actually in the disordered phase. 

\subsection{Absorbing phase transition}
Finally, we discuss the critical behavior of the absorbing phase transition.
Since the symmetry is already broken, it is expected that the model
should belong to the DP class~\cite{DFL2003,HCDM2005}. 
To confirm this, we study the system with the 
initial SM $m_0 = 0.1$. 
In Fig.~\ref{Fig:dp}, we plot $\rho(t) t^{0.45}$ as a function of $t$ on
a semilogarithmic scale, where $0.45$ is the critical exponent of the
DP class. For $\lambda = 0.366\;750$, the curve becomes flat from
around $t = 5\times 10^4$. 
In the active (absorbing) phase, the curves veer up (down) as usual.
Thus we conclude that the critical
point of the absorbing transition is $\lambda_a=0.366\;750(5)$ and the critical behavior
is of the DP class.

Note that exponential decay of $\rho$ in the absorbing phase is observed
in Fig.~\ref{Fig:dp}, which might look inconsistent with 
the power-law decay in the whole absorbing phase reported in
Ref.~\cite{NPKL2011}. However, there is a clear distinction. Since 
the initial SM is nonzero in our case, coarsening has not played any role.
Indeed, we also observe power-law behavior in the absorbing phase 
if $m_0$ is
set to zero just as in Ref.~\cite{NPKL2011} (data not shown). 

\subsection{Summary}
To sum up, we studied a model of two-dimensional interacting monomers,
focusing on the order-disorder phase transition. Numerical analysis
showed that the 2DIM model should belong to the Ising universality class,
contrary to a recent claim~\cite{NPKL2011}.
We observed that analysis of the Binder cumulant is not an efficient
method to find the critical point in two dimensional models with two symmetric 
absorbing states. We also reconfirmed that the absorbing phase transition 
occurring in the 2DIM model after the symmetry is broken 
is described by two-dimensional directed percolation.

Although we did not directly study the interacting monomer-dimer model, 
we believe that the conclusion in this paper
should be applicable to the IMD model studied in Ref.~\cite{NPKL2011}
because of the universality hypothesis.
Since the two different models, IMD and 2DIM, have strong corrections to scaling
at the symmetry-breaking transition point, unlike the Ising model, 
the origin of these strong corrections seems to be 
related to the presence of an absorbing state even for $\lambda>\lambda_c$ 
(disordered phase). If this is the case, it is an interesting question 
as to why and how the absorbing states affect the corrections to scaling;
this is beyond the scope of the present paper and is deferred
to a later presentation.

\begin{acknowledgments}
Discussions with Bongsoo Kim, Sung Jong Lee, and Hyunggyu Park are greatly 
appreciated.  This work was supported 
by the Basic Science Research Program through the National Research Foundation
of Korea (NRF) funded by the Ministry of Education, Science and Technology 
(Grant No. 2010-0006306); and 
by the Catholic University of Korea Research Fund 2011. 
The computation was supported by Universit\"at zu K\"oln, Germany.
\end{acknowledgments}
\bibliographystyle{apsrev}
\bibliography{Park}

\end{document}